# Intrinsic spin polarized electronic structure of $CrO_2$ epitaxial film revealed by bulk-sensitive spin-resolved photoemission spectroscopy

Hirokazu Fujiwara[1], Masanori Sunagawa[1], Kensei Terashima[1,2], Tomoko Kittaka[1], Takanori Wakita[1,2], Yuji Muraoka[1,2], and Takayoshi Yokoya[1,2]

[1]Research Laboratory for Surface Science and Graduate school of Natural Science and Technology, Okayama University, Okayama, 700-8530, Japan

[2]Research Center of New Functional Materials for Energy Production, Storage, and Transport, Okayama University, Okayama, 700-8530, Japan

(Dated: May 18, 2015)

Abstract

We have performed bulk-sensitive spin-resolved photoemission spectroscopy in order to clarify the intrinsic spin-resolved electronic states of half-metallic ferromagnet $CrO_2$. We used $CrO_2$ epitaxial films on $TiO_2$(100), which shows a peak at 1 eV with a clear Fermi edge, consistent with the bulk-sensitive PES spectrum for $CrO_2$. In spin-resolved spectra at 40 K, while the Fermi edge was observed in the spin up (majority spin) state, no states at the Fermi level ($E_F$) with an energy gap of 0.5 eV below $E_F$ was observed in the spin down (minority spin) state. At 300 K, the gap in the spin down state closes. These results are consistent with resistivity measurements and magnetic hysteresis curves of the fabricated $CrO_2$ film, constituting spectroscopic evidence for the half-metallicity of $CrO_2$ at low temperature and reducing the spin polarization at room temperature. We also discuss the electron correlation effects of Cr 3$d$.

**I. INTRODUCTION**

The ferromagnetic metals with an energy gap at the Fermi level ($E_F$) in either one of electronic spin states are called half-metal. The materials are expected to play a central role for spintronics that is a technology using not only the electronic charge degree of freedom but also the electronic spin degree of freedom. Chromium dioxide, $CrO_2$, is a theoretically predicted half-metallic ferromagnet in which the spin polarization at $E_F$ was expected to be 100%.[1-3] The half-metallic behavior of $CrO_2$ was experimentally proved by the point contact Andreev reflection measurement at 1.85 K,[4] where the observed spin polarization at $E_F$ was as high as 96 %. This value has been the highest in the materials determined by the Andreev reflection measurements up to date. Therefore, the half-metallic ferromagnet $CrO_2$ has been regarded as one of the most hopeful materials for spintronics applications.

One of the key ingredients for practical usage is the temperature-dependence of spin polarization, i.e., the device performances at room temperature. Spin-resolved photoemission spectroscopy (spin-resolved PES) is a powerful technique for determining the half-metallicity as a function of temperature. In the earlier studies, the spin polarization of about 100% near $E_F$ was found for the polycrystalline film and film island samples of $CrO_2$ at room temperature by spin-resolved PES using He Iα ($h\nu$ = 21.2 eV) as an excitation light source.[5,6] However, the reported energy positions of Cr 3$d$ band was 2.7 eV (Ref. 5) and 2.3 eV (Ref. 6), away from the known value of ~ 1 eV which was reported in spin-integrated PES by other groups.[7,8] In addition, the photoelectron intensity at $E_F$ was too small to be regarded as a metal. It is well known that the surface of $CrO_2$ easily transforms into antiferromagnetic insulator $Cr_2O_3$,[9] whose energy position in photoemission spectra was reported to be ~ 2 eV.[10,11] In the previous spin-resolved PES study,[6] the maximum spin polarization were observed after removing the surface layer by moderate sputtering for certain optimum periods and further sputtering reduced the spin polarization. Thus it is likely that the $Cr_2O_3$ formed on the surface and sputtering procedure significantly influenced the results of the previous spin-resolved PES studies because of the surface sensitivity of the detected photoelectrons excited by He Iα. To determine the intrinsic spin-dependent electronic structure and discuss the origin of half-metallicity, bulk-sensitive study has been anticipated. The escape depth of photoelectrons in a solid depends on their kinetic energies[12] and the use of Xe I line ($hv$ = 8.44 eV) enable us to perform bulk-sensitive measurements,[13] since the escape depth of excited photoelectrons is relatively long (~ 50 Å[12]).

In this letter, we present a bulk-sensitive spin-resolved PES study of high quality $CrO_2$ films[7] at 40 K and 300 K. We have observed the intrinsic electronic states of $CrO_2$ peaked at 1.0 eV in the spin up spectra with a clear Fermi edge and the peak top of 1.0 eV in the spin up spectra at both 40 K and 300 K. A band gap of 0.5 eV below $E_F$ was observed in the spin down state at 40 K, which supports the half-metallicity of $CrO_2$ at low temperature, whose magnitude suggests that the electron correlation $U$ firmly affects to the electronic states.

## II. EXPERIMENTAL METHODS

The $CrO_2$(100) films was prepared on $TiO_2$(100) substrates with the same chemical vapor deposition (CVD) technique as what reported earlier by Iwai *et al.*[7] The thickness of the films was estimated to be approximately 100 nm. After the synthesis, the $CrO_2$ film was taken from the quartz tube in a helium atmosphere and then immediately introduced into the ultra-high vacuum (UHV) for the spin-resolved PES. The quality of prepared $CrO_2$ film surface was evaluated just before the PES measurements by low energy electron diffraction (LEED), as shown in Fig. 1(a). The rectangular-like pattern characteristic of tetragonal crystal structure[7] was confirmed on the sample surface without any cleaning procedures including sputtering and annealing. The LEED pattern is consistent with that of $CrO_2$(100) reported in previous studies.[14]

The spin-resolved PES measurements were carried out at 40 K and 300 K in a spin-resolved PES system with a base pressure of 1 × 10$^{-8}$ Pa at Okayama University. The unpolarized light of He I line and Xe I line were used to excite photoelectrons for the experiments. The light of Xe Iα line was monochromated by a $CaF_2$ filter. The energy resolution, during the spin-integrated and spin-resolved PES, was approximately 35 meV and 100 meV, respectively. The spin-integrated and spin-resolved PES spectra were obtained in the transmission mode of the detector. The acceptance angle of the analyzer was ± 15˚ along [001] direction (easy axis) and ± 1˚ along [010] direction, corresponding to the ~ 50 % Brillouin zone along [001] direction in the ΓXRZ plane centered at the Γ(X) point ( $k_z$ is unknown at present ). To resolve the direction of the spin of photoelectrons, we used a mini Mott spin detector (VG Scienta 2D spin)[15] whose target was a Au polycrystalline film. The effective Sherman function $S_{eff}$ was determined to be 0.1 by spin- and angle-resolved PES measurements of Bi/Si(111) thin film. The sensitivity of a pair of detectors in the Mott polarimeter was calibrated by non-polarized photoelectrons from Au polycrystalline film. We magnetized the sample along the easy axis by bringing a magnet close to the sample and measured spin polarization along the magnetized direction. Careful attention has been paid to hold the direction of magnetization when we took the magnet away from the sample.

## III. RESULTS AND DISCUSSION

Figure 1(b) shows spin-integrated valence band PES spectra near $E_F$ of epitaxial $CrO_2$(100) film by the He I and the Xe I lines at 300 K. In the spectrum of the He I line (He I spectrum), the peak positions were found to be approximately 1 eV and 2 eV. The intrinsic peak position derived from $CrO_2$ was reported to be at 1.0 eV by bulk sensitive hard x-ray photoemission spectroscopy (HAXPES),[8] Therefore we identify the peak of approximately 1 eV as Cr 3$d$ states of $CrO_2$ hybridized with O 2$p$ states. Existence of a structure approximately 2 eV indicates that small amount of $Cr_2O_3$ exists on the $CrO_2$ film, even if the only $CrO_2$ derived LEED pattern for the sample was observed. The relative intensity of 2 eV structure depends on films and for spin-resolved PES measurements we used a film with negligible 2 eV structure. A clear Fermi edge in addition to the 1 eV structure was also observed in the He I spectrum, which indicates that the quality of samples used in the present study were higher than those of the previous studies.[5,6] In a previous resonant PES study of $CrO_2$,[14] the peak position of Cr 3$d$ was found to be 1.9 eV with the LEED pattern of $CrO_2$ without $Cr_2O_3$-derived spots, similar to our measurements of the He I. In the Xe I spectrum, the spectral weight around 1 eV is relatively enhanced as compared to the He I spectrum, which implies the energy distribution curve (EDC) of Xe I reflects more intrinsic electronic states of $CrO_2$ than that of the He I. Next we discuss the spin polarization of the bulk electronic states of $CrO_2$ using the spin-resolved Xe I spectra.

Figures 1 (c) and (d) show the spin-resolved EDCs of the $CrO_2$ film and the spin polarization as a function of binding energy at 300K and 40 K, respectively. The spin polarization from $E_F$ to ~ 2 eV was observed both at 40 K and 300 K. It is clear from the figure that the spin down spectrum shows the insulating energy gap of 0.5 eV at least in the occupied side at 40 K (green thick line in Fig. 1(d)), while the gap in the spin down spectrum at 300 K is closed. The degrees of spin polarization were 40 % and 100 % near $E_F$ at 300 K and 40 K, respectively, which is consistent with magnetic hysteresis curves.[7] These results are different from the previous spin-resolved PES study[6] in three points: (i) The spin-polarized binding energy region ($E_F$ ~ 2 eV) is narrower than the one in the previous study ($E_F$ ~ 6 eV) and more consistent with the band calculation predictions. (ii) The spin polarization near $E_F$ at room temperature of 40 % is smaller than the one of 95 % in the previous study and closer to the values expected from magnetization hysteresis curves.[7] (iii) The clear Fermi edge is observed without any sputtering while the intensities of the Fermi edge in the previous studies are too small to be regarded as a metal even after the optimized sputtering procedures. Although we do not have a direct answer for the cause of these differences, we note that the experimental conditions, specifically the excitation photon energy and the conditions of the film formation are different. Since we performed more bulk-sensitive measurements on high quality samples than the previous studies, we believe that the present results reflect the nature of intrinsic $CrO_2$ film with less influence of $Cr_2O_3$. One of the previous studies argued that the highly spin polarized results of the previous spin-resolved PES studies at room temperature might have suffered the $Cr_2O_3$ surface layer in a magnetically ordered single domain state on top of the $CrO_2$ film.[16] Our LEED patterns before and after spin-resolved PES measurements did not show the six-fold pattern derived from ordered $Cr_2O_3$, but with relatively high background, indicating that the $Cr_2O_3$ layer was in an amorphous state. The spin-polarized energy region corresponds to that of the coherent structure around 1 eV of HAXPES spectra,[8] suggesting that the intrinsic electronic structure of $CrO_2$ was obtained. The consistency between the present spin-resolved PES spectra and the macroscopic measurements,[7] i.e. magnetic hysteresis curves and resistivity measurements, suggests that the effects of $Cr_2O_3$ on the $CrO_2$ film is small and supports the reliability of our spin-resolved PES results.

Next we discuss the possible reasons of depolarization near $E_F$ at 300 K. As seen in Figs. 1(c) and 1(d), the intensity between 2 eV and $E_F$ in the spin down states increases with increasing temperature, which primarily causes the decrease of spin polarization at 300 K. As the reasons why the spin polarization near $E_F$ decreases at 300 K, the change of the amount of exchange splitting, formation of nonquasiparticle (NQP) states,[17] or formation of magnetic domains are considered. In general, the energy scale of the change of the amount of exchange splitting below the Currie temperature ($0.3 < T / T_C < 0.8$) is 10 – 100 meV,[18,19] which is different from the energy region that the spin polarization decreases. The NQP states in $CrO_2$ were predicted by Chioncel et al[20] who suggested that the NQP states can be produced near and above $E_F$. In this case, the down spin states in the vicinity of $E_F$ would increase rather than those at the higher binding energy part, which contradicts to what observed here. In addition, the energy scale of NQP states below $E_F$ has been argued to be about 100 meV which does not explain the observed spin depolarization over a larger energy scale. Therefore we concluded that the decrease of spin polarization at 300 K is mainly due to formation of magnetic domains. This picture is supported by the fact that the energy region where the spectral intensity of spin down states increases with increasing temperature matches well with the energy region where spin up states populate. The above discussion suggests that higher spin polarizations in $CrO_2$ at room temperature can be obtained by control of magnetic domains.

Finally, we discuss the on-site Coulomb interaction by comparison of experimental spin-resolved EDCs at 40 K with the density of states calculated by the LSDA+$U$ method using 0, 3, and 6 eV as the electron correlation $U$,[3] as shown in Fig. 2. The spectral characteristics of spin-resolved PES, e.g. the peak position and the intensity at $E_F$ of the spin up states and the gap on the occupied side of the spin down states, fits more closely with the band calculations using $U$ = 3 eV or 6 eV rather than that of $U$ = 0 eV, indicating that electron correlation $U$ must have a finite value in $CrO_2$. Along with the previous hard x-ray PES and inverse PES study[8] which discuss the value of $U$, our spin-resolved PES study points out that electron correlation $U$ must have been taken into account when considering the electronic structure of half-metallic ferromagnet $CrO_2$ system.

## IV. CONCLUSION

We have investigated the temperature dependence of the spin-dependent electronic structure of high quality $CrO_2$(100) epitaxial film by bulk-sensitive spin-resolved PES. A clear Fermi edge and spin polarization of about 100 %, and a band gap of 0.5 eV in the spin down states were observed at 40 K, providing the spectroscopic evidence for the half-metallicity of $CrO_2$ at low temperatures, at least in the probed momentum region of the Brillouin zone. In addition, the Fermi edge and spin polarization of 40 % near $E_F$ was observed at 300 K, which is consistent with earlier reports of resistivity measurements and magnetic hysteresis curves. The comparison of the spin-resolved PES spectra with LSDA+$U$ band calculations suggests that the electronic correlation $U$ must be considered in order to discuss the electronic states of $CrO_2$. These results also indicate that bulk-sensitive spin-resolved PES is a very powerful experimental tool to investigate intrinsic spin-resolved electronic states of half-metallic ferromagnets.

## ACKNOWLEDGEMENTS

We acknowledge M. Ogata for technical assistance. We thank T. Okuda, K. Miyamoto, and T. Kinoshita for the valuable discussions. We also thank A. Ino for his ARPES data analysis program. This work was partially supported by the Program for Promoting the Enhancement of Research Universities and a Grant-in-Aid for Young Scientists (B) (No.25800205) from the Ministry of Education, Culture, Sports, Science and Technology of Japan (MEXT).

[§]E-mail: sc422226@s.okayama-u.ac.jp

Figure captions

FIG. 1. (Color online) (a) LEED pattern of the $CrO_2$(100) epitaxial film on $TiO_2$. The energy of incident electrons was set at 45 eV. (b) Valence band spin-integrated PES spectra measured by the He I line (21.2 eV) and a Xe I line (8. 44 eV) at 300 K. (c), (d) Spin-resolved EDCs measured by the Xe I line and corresponding energy dependence of the spin polarization along the easy axis direction at 300 K and 40 K, respectively.

FIG. 2. (Color online) Comparison of the experimental spin-resolved EDCs at 40 K in the present study with theoretical spin-resolved density of states calculated by Jeng and Guo.[3] These band calculations are multiplied by the Fermi-Dirac function and broadened by convoluting with a 100 meV Gaussian.

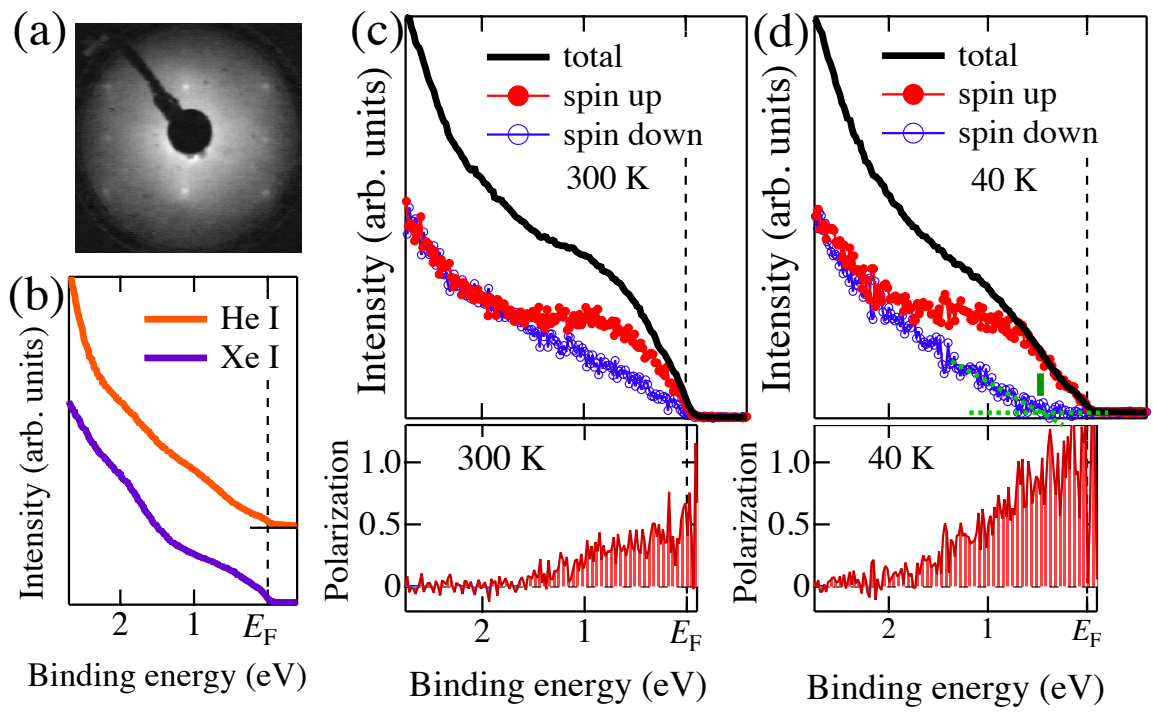

FIG. 1   H. Fujiwara

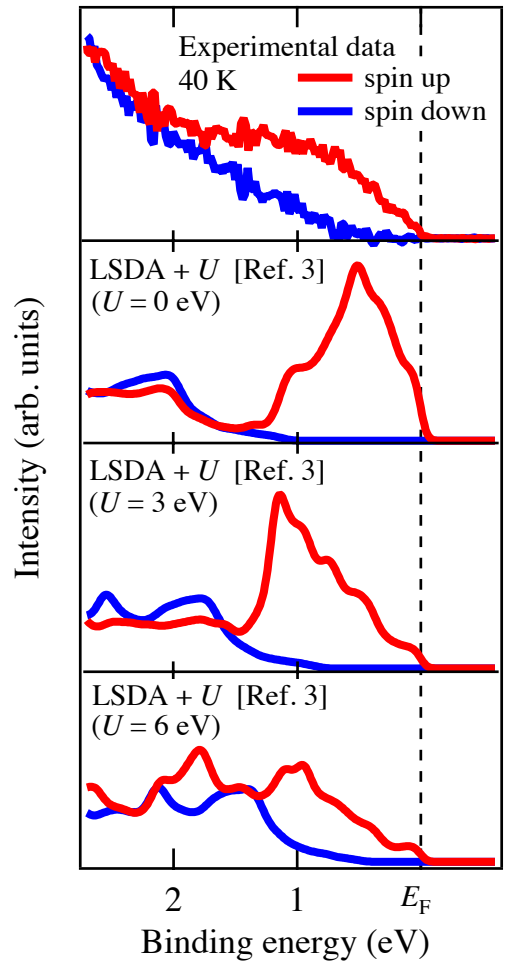

FIG. 2  H. Fujiwara